\documentclass[aps,groupedaddress,showpacs,floatfix]{revtex4}
\usepackage{amssymb}
\usepackage{amsmath}
\usepackage{graphicx}
\begin{document}

\thispagestyle{empty}

\title{On the scaling properties of (2+1) directed polymers in the high temperature limit}

\author{Victor Dotsenko$^{\, a,b}$ and Boris Klumov$^{\, c}$}

\affiliation{$^a$LPTMC, Sorbonne Universit\'e, Paris, France}

\affiliation{$^b$Landau Institute for Theoretical Physics,
   Moscow, Russia}

\affiliation{$^c$Joint Institute for High Temperatures, Russian Academy of Sciences,
   Moscow, Russia}

\date{\today}

\begin{abstract}
In this paper in terms of the replica method we consider the high temperature limit 
of (2+1) directed polymers in a random potential and 
propose an approach which allows to compute  
the scaling exponent $\theta$ of the free energy fluctuations as well as the 
left tail of its probability distribution function. It is argued that 
$\theta = 1/2$ which is {\it different} from the zero-temperature numerical
value which is close to $0.241$. 
This result implies that unlike the $(1+1)$ system in the two-dimensional case
the free energy scaling exponent is non-universal being temperature dependent.
\end{abstract}


\maketitle

\section{Introduction}

At present the statistical properties of one-dimensional directed polymers
as well as the other systems belonging to the so called KPZ universality class
\cite{KPZ} are sufficiently well studied 
(for the reviews see e.g. \cite{Corwin,Borodin,Rev}).
In contrast to that, not so much is achieved in the studies of the 
so called (2+1) model of directed polymers which describes the fluctuations 
of an elastic string directed along the time axes which passes through 
a random medium in the three-dimensional space. 
Due to extensive numerical simulations it is rather convincingly 
established that at  the zero-temperature and  in the limit of 
large times $t$ the free energy fluctuations  
of such directed polymers scale as $ t^{\theta}$ with the scaling exponent 
$\theta \simeq 0.241$ \cite{numerics1,numerics2,numerics3}.
The third digit here is still disputed but there is unanimous agreement that
$\theta$  is {\it not equal} to $1/4$ (as every theoretician would hope).
On the other hand, on the theoretical side, there is a rigorous
proof that in the {\it high temperature limit} the disorder potential
remains relevant \cite{Lacoin,Berger-Lacoin,Comets}. In other words, there is 	effective localization 
of the polymers trajectories around disorder defined favorite 
"corridors" so that (like in the (1+1) system) even at high temperatures 
the polymers statistics do not reduce to simple diffusion 
(as it takes place e.g. in the high temperature phase 
of the (3+1) model\cite{Bolthausen}).

In this paper in terms of the replica technique we propose an approximate
(mean field) method which in the {\it high temperature limit} allows to estimate
the replica partition function in the limit of {\it large number} of 
replicas. This in turn, makes possible to derive the {\it left tail} asymptotics
of the free energy fluctuations distribution function. Assuming that this 
(unknown) distribution function is defined by only one energy scale 
one eventually finds that the typical value of the disorder defined free energy fluctuations
scale as $ \sim \sqrt{u} \, t^{1/2}$ (where $u$ is the parameter 
which describes the strength of the disorder) which means that the time 
scaling exponent $\theta=1/2$. 
This value is remarkably  different from the 
zero temperature numerical results. If correct, this statement implies 
that unlike the $(1+1)$ system (where it is rigorously proved that $\theta=1/3$ 
both at $T\to 0$ and at $T\to\infty$) in the two-dimensional case 
the free energy scaling exponent is non-universal being temperature dependent. 
Note also that with $\theta = 1/2$ for the mean square fluctuations of the polymer trajectory $\boldsymbol{\phi}(\tau) \; (0 \leq \tau \leq t)$ 
simple energy balance arguments  yields the scaling 
$\overline{\phi^{2}} \sim t^{\theta+1} = t^{3/2}$.

It is  also worthing to note that comparing the features of the high temperature 
limits in	(1+1) and (2+1) systems one finds one more interesting point. 
In (1+1) dimension there is the high temperature
regime in which the effective temperature parameter is rescaled with time 
as $\beta \to \beta \, t^{-1/4}$ and intermediate exponents are established 
\cite{Albert-Khanin-Quastel-1,Albert-Khanin-Quastel-2}.
In particular, in the limit $t \to \infty$ this scaling interpolates between 
the weak disorder ($\beta=0$) and strong disorder regimes ($\beta > 0$). 
In a sense, this result shares some features of the problem considered  
in the present paper.

In Section II we define the  model and describe the general ideas
of the present approach. In Section III we present the systematic mean-field method
(hopefully valid in the limit of large number of replicas) 
which eventually can be reduced to the  solution of two-dimensional 
one-particle non-local (integral)  {\it non-linear }
differential equation (\ref{32}). In the high temperature limit
this equation can be (numerically) solved providing the value of 
the exponent $\theta=1/2$ (Section IV). 
Section V is devoted to the brief discussion of the obtained results.

\section{The model and the replicas approach} 

We consider the model of directed polymers defined in terms
of an elastic string described by the two-dimensional vector
$\boldsymbol{\phi}(\tau) \equiv \bigl(\phi_{x}(\tau), \, \phi_{y}(\tau)\bigr)$ 
directed along the $\tau$-axes within an interval $[0,t]$ 
which passes through a random medium
described by a random potential $V(\boldsymbol{\phi},\tau)$. 
The energy of a given polymer's trajectory
$\boldsymbol{\phi}(\tau)$ is
\begin{equation}
\label{1}
H[\boldsymbol{\phi}(\tau); V] = \int_{0}^{t} d\tau
\biggl\{\frac{1}{2} \bigl[\partial_\tau \boldsymbol{\phi}(\tau)\bigr]^2
+ V[\boldsymbol{\phi}(\tau),\tau]\biggr\};
\end{equation}
Here the disorder potential $V[\boldsymbol{\phi},\tau]$
is supposed to be Gaussian distributed with a zero mean
$\overline{V(\boldsymbol{\phi},\tau)}=0$
and the correlation function
\begin{equation}
\label{2}
\overline{V(\boldsymbol{\phi},\tau)V(\boldsymbol{\phi}',\tau')} = 
u \, \delta(\tau-\tau') U(\boldsymbol{\phi}-\boldsymbol{\phi}')
\end{equation}
The parameter $u$ is the strength of the disorder and $U(\boldsymbol{\phi})$ is a smooth
function characterized by the correlation length $\epsilon$. For simplicity we take
\begin{equation}
\label{3}
U(\boldsymbol{\phi}) \; = \; \frac{1}{2\pi \, \epsilon^{2}} \; 
            \exp\Bigl\{-\frac{\boldsymbol{\phi}^{2}}{2 \epsilon^{2}}\Bigr\}
\end{equation}

One-dimensional, or the so called (1+1) version of this problem (when instead of the
vector we have a scalar field $\phi(\tau)$)  with the $\delta$-correlated random potential has been the focus  of intense studies during past three
decades
\cite{hhf_85,numer1,numer2,kardar_87,bouchaud-orland,hh_zhang_95,
	Johansson,Prahofer-Spohn,Ferrari-Spohn1,KPZ-TW1a,KPZ-TW1b,KPZ-TW1c,KPZ-TW2,BA-TW2,BA-TW3,
	LeDoussal1,LeDoussal2,goe,end-point,LeDoussal3}.
At present it is well established that the fluctuations of the free energy
of this system 
are described by the Tracy-Widom  (TW) distribution \cite{TW-GUE}
and their typical value scale with time as $t^{1/3}$.

%%%%%%%%%%%%%%%%%%%%%%%%%%%%%%%%%%%%%%%

\vspace{5mm}

The general formulation of the considered  (2+1) problem in terms of 
the replica approach looks quite similar to the (1+1) one.
For a given realization of the random potential $ V(\boldsymbol{\phi},\tau)$
the partition function of the considered system (with fixed boundary conditions) is
\begin{equation}
\label{4}
Z({\bf r}, t) = \int_{\boldsymbol{\phi}(0)=\bf{0}}^{\boldsymbol{\phi}(t)={\bf r}} 
{\cal D}\boldsymbol{\phi}(\tau)
\exp\bigl\{-\beta H[\boldsymbol{\phi}(\tau), V]\bigr\} \; = \; \exp\bigl\{-\beta F({\bf r}, t)\bigr\}
\end{equation}
where $\beta$ is the inverse temperature,
$F({\bf r},t)$ is the free energy which is a random quantity 
and the integration is taken over all
trajectories $\boldsymbol{\phi}(\tau)$
starting at $\bf{0}$ (at $\tau = 0$) and ending at a point ${\bf r}$ 
(at $\tau = t$). 
Note that this problem is equivalent to the KPZ equation \cite{KPZ}
\begin{equation}
\label{4a}
\partial_{t} F({\bf r}, t) \; = \; \frac{1}{2\beta} \nabla^{2} F({\bf r}, t) 
\; - \; \frac{1}{2} \Bigl(\boldsymbol{\nabla} \, F({\bf r}, t)\Bigr)^{2} \; + \; 
 V({\bf r}, t)
\end{equation}
which describe the time evolution of the two-dimensional manifold $F({\bf r}, t)$
in a random potential $V({\bf r}, t)$.

For simplicity, in what follows we are going to consider the problem with
the zero boundary conditions: $\boldsymbol{\phi}(\tau=0)=\boldsymbol{\phi}(\tau=t)={\bf 0}$.
The free energy probability distribution function
$P(F)$ of this system can be studied in terms of the integer moments 
of the above partition function, eq.(\ref{4}):
\begin{equation}
\label{5}
\overline{Z^{N}} \equiv Z(N,t) \; = \; 
\prod_{a=1}^{N}\int_{\boldsymbol{\phi}_{a}(0)=0}^{\boldsymbol{\phi}_{a}(t)=0}
{\cal D}\boldsymbol{\phi}_{a}(\tau) \;
\overline{
	\Biggl(
	\exp\Bigl\{-\beta \sum_{a=1}^{N} H[\boldsymbol{\phi}_{a}(\tau), V]\Bigr\}
	\Biggr)}
\; = \;
\int_{-\infty}^{+\infty} dF \, P(F) \, \exp\bigl\{-\beta N F\bigr\}
\end{equation}
where $\overline{(...)}$ denotes the averaging over 
the random potentials $V[\boldsymbol{\phi},\tau]$ 
Performing this simple Gaussian averaging we get
\begin{equation}
\label{6}
Z(N,t) \; = \; \prod_{a=1}^{N}\int_{\boldsymbol{\phi}_{a}(0)=0}^{\boldsymbol{\phi}_{a}(t)=0}
{\cal D}\boldsymbol{\phi}_{a}(\tau) \;
\exp\Bigl\{-\beta  
H_{N}[\boldsymbol{\phi}_{1}(\tau), \, \boldsymbol{\phi}_{2}(\tau), \, ... \, , \boldsymbol{\phi}_{N}(\tau)]
\Bigr\}
\end{equation}
where
\begin{equation}
\label{7}
\beta H_{N}[\boldsymbol{\phi}_{1}(\tau), \, \boldsymbol{\phi}_{2}(\tau), \, ... \, , \boldsymbol{\phi}_{N}(\tau)] =  \int_{0}^{t} d\tau
\Biggl[
\frac{1}{2} \beta \sum_{a=1}^{N} \Bigl(\partial_\tau \boldsymbol{\phi}_{a}(\tau)\Bigr)^2
\; - \;
\frac{1}{2} \beta^{2} \, u \,
\sum_{a,b=1}^{N} U\bigl(\boldsymbol{\phi}_{a}(\tau) - \boldsymbol{\phi}_{b}(\tau)\bigr)
\Biggr];
\end{equation}
is the replica Hamiltonian which describes $N$ elastic strings
$\bigl\{\boldsymbol{\phi}_{1}(\tau), \, \boldsymbol{\phi}_{2}(\tau), \, ... \, , \boldsymbol{\phi}_{N}(\tau)\bigr\}$
with the attractive interactions $U\bigl(\boldsymbol{\phi}_{a} - \boldsymbol{\phi}_{b}\bigr)$, eq.(\ref{3}).
To compute the replica partition function $Z(N,t)$, eq.(\ref{6}), one introduces
the function:
\begin{equation}
\label{8}
\Psi({\bf r}_{1}, \, {\bf r}_{2}, \, ... \, {\bf r}_{N}; \; t) \; = \;
\prod_{a=1}^{N}\int_{\boldsymbol{\phi}_{a}(0)=\bf{0}}^{\boldsymbol{\phi}_{a}(t)={\bf r}_{a}}
{\cal D}\boldsymbol{\phi}_{a}(\tau) \;
\exp\Bigl\{-\beta  H_{N}[\boldsymbol{\phi}_{1}(\tau), \, \boldsymbol{\phi}_{2}(\tau), \, ... \, , \boldsymbol{\phi}_{N}(\tau)]\Bigr\}
\end{equation}
such that 
\begin{equation}
\label{8a}
Z(N,t) \, = \, \Psi({\bf r}_{1}, \, {\bf r}_{2}, \, ... \, {\bf r}_{N}; \, t)\big|_{{\bf r}_{a}=0}
\end{equation}
Here the spatial arguments of this function are $N$
two-dimensional vectors $\{{\bf r}_{a}\}$.
One can easily show that $\Psi({\bf r}_{1}, \, {\bf r}_{2}, \, ... \, {\bf r}_{N}; \, t)$ is the wave function of N quantum bosons defined by
the imaginary time Schr\"odinger equation
\begin{equation}
\label{9}
\beta \frac{\partial}{\partial t} 
\Psi({\bf r}_{1}, \, {\bf r}_{2}, \, ... \, {\bf r}_{N}; \; t) \; = \;
\frac{1}{2}\sum_{a=1}^{N} \, \Delta_{a} 
\Psi({\bf r}_{1}, \, {\bf r}_{2}, \, ... \, {\bf r}_{N}; \; t)
\; + \; \frac{1}{2} \, \beta^{3} u \, \sum_{a,b=1}^{N} U({\bf r}_{a} - {\bf r}_{b}) \,
\Psi({\bf r}_{1}, \, {\bf r}_{2}, \, ... \, {\bf r}_{N}; \; t)
\end{equation}
where $\Delta_{a}$ is the two-dimensional Laplacian 
with respect to the coordinate ${\bf r}_{a}$.  The corresponding eigenvalue equation 
for the eigenfunctions $\psi({\bf r}_{1}, \, {\bf r}_{2}, \, ... \, {\bf r}_{N})$, 
defined by the relation
\begin{equation}
\label{10}
\Psi({\bf r}_{1}, \, {\bf r}_{2}, \, ... \, {\bf r}_{N}; \; t) \; = \;
\psi({\bf r}_{1}, \, {\bf r}_{2}, \, ... \, {\bf r}_{N}) \, 
\exp\bigl\{-t \, E_{N}\bigr\}
\end{equation}
reads:
\begin{equation}
\label{11}
-2 \beta \, E_{N} \, 
\psi({\bf r}_{1}, \, {\bf r}_{2}, \, ... \, {\bf r}_{N}) \; = \;
\sum_{a=1}^{N} \, \Delta_{a} 
\psi({\bf r}_{1}, \, {\bf r}_{2}, \, ... \, {\bf r}_{N})
\; + \; \beta^{3} u \, \sum_{a,b=1}^{N} U({\bf r}_{a} - {\bf r}_{b}) \,
\psi({\bf r}_{1}, \, {\bf r}_{2}, \, ... \, {\bf r}_{N})
\end{equation}

It is at this stage that we are facing the crucial difference of the considered problem
with the corresponding (1+1) one. The general solution
of the one-dimensional counterpart of eq.(\ref{11}) is given by the Bethe ansatz 
wave function which is valid only for $U(x) \, = \, \delta(x)$ and which is based on the 
exact two-particle wave functions ($N=2$) solution 
exhibiting  finite value  energy $E_{N=2}$. It is this fundamental property
of (1+1) problem which eventually allows to derive the Tracy-Widom distribution
for the free energy fluctuation. 

The situation in the (2+1) case, eq.(\ref{11}), is much more complicated. 
First of all, in two dimensions there exists no {\it finite} two-particle solution
for $U({\bf r}) \, = \, \delta({\bf r})$. One can easily construct an approximate 
ground state solution of the two-particle problem for the finite-size function $U({\bf r})$,
eq.(\ref{3}), but then one finds that in the limit $\epsilon \to 0$ (when $U({\bf r})$
turns into the $\delta$-function) the ground state energy of this solution 
$E_{N=2} \, \to \, -\infty$. In other words, in two dimensions (unlike one-dimensional case)
one can not consider the problem with $\delta$-correlated random potential. 
We have to study the system with {\it finite} size function $U({\bf r})$ and 
the value of its spatial size $\epsilon$ must explicitly enter into the final 
results.

Second, one can easily demonstrate that in the two-dimensional case 
the construction of the $N$-particle 
wave function {\it \`a la} Bethe ansatz structure based on the approximate two-particle
solution (for finite $\epsilon$) doesn't work. So that, unlike one-dimensional case, 
here even the ground state energy $E_{N}$ as well as $N$-particle 
ground state wave function 
$\psi({\bf r}_{1}, \, {\bf r}_{2}, \, ... \, {\bf r}_{N})$ of eq.(\ref{11})
are not known. All that makes the perspective to find the exact solution of the (2+1)
problem rather doubtful.

%%%%%%%%%%%%%%%%%%%%%%%%%%%%%%%%%%%%%%

Here we would like to propose somewhat different strategy which, 
at least in the high-temperature limit, makes
possible to estimate the replica partition function $Z(N, t)$ at $N \gg 1$
which in turn allows to derive the scaling exponent of the free energy fluctuations.

\section{Mean Field Approach}

First, to simplify notations, let us eliminate
the parameter $\epsilon$ of the correlation function $U\bigl({\bf r}\bigr)$.
Redefining 
\begin{equation}
\label{22a}
{\bf r} \; \to \; \epsilon \, {\bf r}
\end{equation}
and 
\begin{equation}
\label{22}
E_{N} \; = \;  \frac{1}{\epsilon^{2}} \, {\tilde E}_{N} \, ,
\end{equation}
instead of eq.(\ref{11}) we get
\begin{equation}
\label{23}
-2 \beta \, \tilde{E}_{N} \, 
\psi({\bf r}_{1}, \, {\bf r}_{2}, \, ... \, {\bf r}_{N}) \; = \;
\sum_{a=1}^{N} \, \Delta_{a} 
\psi({\bf r}_{1}, \, {\bf r}_{2}, \, ... \, {\bf r}_{N})
\; + \; \beta^{3} u \, \sum_{a,b=1}^{N} U_{0}({\bf r}_{a} - {\bf r}_{b}) \,
\psi({\bf r}_{1}, \, {\bf r}_{2}, \, ... \, {\bf r}_{N})
\end{equation}
where
\begin{equation}
\label{24}
U_{0}({\bf r}) \; = \; \frac{1}{2\pi} \; 
 \exp\biggl\{-\frac{1}{2} \, {\bf r}^{2} \biggr\}
\end{equation}
%

%%%%%%%%%%%%%%%%%%%%%%%%%%%%%%%%%%%%%%%%%%%%%

\vspace{10mm}

It is evident that in a general case eq.(\ref{23}) can not be solved.
However in the limit of large number of particles, $N \gg 1$, 
one hopefully can use the standard trick of the mean field approximation,
in which the $N$-particle wave function 
$\psi({\bf r}_{1}, \, {\bf r}_{2}, \, ... \, {\bf r}_{N})$ factorizes into the 
product of $N$ one-particle functions, namely
\begin{equation}
\label{25}
\psi({\bf r}_{1}, \, {\bf r}_{2}, \, ... \, {\bf r}_{N}) \; \simeq \; 
\prod_{a=1}^{N} \psi({\bf r}_{a})
\end{equation}
Substituting this into eq.(\ref{23}), we obtain
\begin{equation}
\label{26}
-2 \beta \, \tilde{E}_{N} \,\prod_{a=1}^{N} \psi({\bf r}_{a}) \; = \;
\sum_{a=1}^{N} \, \Delta_{a} \psi({\bf r}_{a}) \; 
\prod_{b\not=a}^{N} \psi({\bf r}_{b})
\; + \; \beta^{3} u \, \sum_{a\not=b}^{N} U_{0}({\bf r}_{a} - {\bf r}_{b}) \,
\psi({\bf r}_{a}) \psi({\bf r}_{b}) \;
\prod_{c\not=a,b}^{N} \psi({\bf r}_{c}) \, + \,
\frac{1}{2\pi}\beta^{3} u N \prod_{a=1}^{N} \psi({\bf r}_{a})
\end{equation}
Introducing notations
\begin{eqnarray}
\label{27}
\tilde{E}_{N}      &=& -\frac{N}{4\beta} \, \lambda
\\
\nonumber
\\
\label{28}
\beta^{3} u \, N &=& \frac{1}{2} \kappa
\end{eqnarray}
and integrating eq.(\ref{26}) over all 
${\bf r}_{1}, \, {\bf r}_{2}, \, ... \, {\bf r}_{N}$
(taking into account that $\int d^{2}r \Delta \psi({\bf r}) =0$)
we get
\begin{equation}
\label{29}
\kappa (N-1) C_{0}^{-2} \int d^{2}r_{1} \int d^{2}r_{2} 
U_{0}({\bf r}_{1} - {\bf r}_{2}) \,
\psi({\bf r}_{1}) \psi({\bf r}_{2}) \; = \; 
\lambda \, N \; - \; \frac{1}{2\pi} \kappa
\end{equation}
where $C_{0} \; = \; \int d^{2}r \, \psi({\bf r})$.
Now, integrating eq.(\ref{26}) over 
${\bf r}_{2},  \, ... \, ,{\bf r}_{N}$, using the above relation,
eq.(\ref{29}), redefining $\psi({\bf r}) \to  C_{0} \, \psi({\bf r})$
and neglecting terms of order $N^{-1}$ we get the following non-linear
mean-field equation for the one-particle function $\psi({\bf r})$:
\begin{equation}
\label{32}
\Delta \psi({\bf r}) \; - \; 
\lambda \, \psi({\bf r}) \; + \; 
\kappa \, \psi({\bf r}) \,
\int d^{2} r' \, U_{0}({\bf r} - {\bf r}') \, \psi({\bf r}') \; = \; 0
\end{equation}
where  
\begin{equation}
\label{33}
\int d^{2}r \, \psi({\bf r}) \; = \; 1
\end{equation}
and the function $U_{0}({\bf r})$ is given in eq.(\ref{24}). 

Further strategy is the following. 
For given values of the parameters $\lambda$ and $\kappa$,  eq.(\ref{27}),
we have to find 
smooth non-negative solution of eq.(\ref{32}) such that 
$\psi({\bf r}\to\infty) \to 0$.
Next, substituting this solution into the constraint (\ref{33}) we can find
$\lambda$ as a function of $\kappa$, which eventually gives us the dependence of
the ground state energy, eqs.(\ref{22}) and (\ref{27}), on the replica parameter $N$.
First, let us demonstrate how this strategy works in the well studied 
one-dimensional case.

\subsection{The example of $(1+1)$ system}

The one-dimensional version of eqs.(\ref{32})-(\ref{33}) reads
\begin{equation}
\label{34}
\psi''(x) \; - \; \lambda \, \psi(x) \; + \; 
\kappa \, \psi(x) \,\int d x' \, U_{1}(x - x') \, \psi(x') \; = \; 0
\end{equation}
where 
\begin{equation}
\label{35}
\int_{-\infty}^{+\infty} dx \, \psi(x) \; = \; 1
\end{equation}
and 
\begin{equation}
\label{36}
U_{1}(x) \; = \; \frac{1}{\sqrt{2\pi}} \; 
\exp\biggl\{-\frac{1}{2} \, x^{2} \biggr\}
\end{equation}
Redefining
\begin{equation}
\label{37}
\psi(x) \; = \; \frac{\lambda}{\kappa} \, \phi\bigl(\sqrt{\lambda} \, x)
\end{equation}
and denoting $\sqrt{\lambda} \, x \; = \; z$,
instead of eqs.(\ref{34})-(\ref{35}) we get
\begin{equation}
\label{38}
\phi''(z) \; - \;  \phi(z) \; + \;  
\phi(z)\, \int d z' \, U_{\lambda}(z - z') \, \phi(z') \; = \; 0 
\end{equation}
where the function $\phi(z)$ satisfy the constraint
\begin{equation}
\label{39}
\frac{\sqrt{\lambda}}{\kappa} \, \int_{-\infty}^{+\infty} dz \, \phi(z) \; = \; 1
\end{equation}
and
\begin{equation}
\label{40}
U_{\lambda}(z) \; = \; \frac{1}{\sqrt{2\pi\lambda}} \; 
\exp\biggl\{-\frac{1}{2\lambda} \, x^{2} \biggr\}
\end{equation}
According to eq.(\ref{39}),
\begin{equation}
\label{41}
\lambda(\kappa) \; = \; I^{-2} \, \kappa^{2}
\end{equation}
where
\begin{equation}
\label{42}
I \; = \;  \int_{-\infty}^{+\infty} dz \, \phi(z)
\end{equation}
According to eq.(\ref{41}) in the high temperature limit both  
$\kappa \propto \beta^{3} u N \to  0$ and $\lambda \to 0$. 
Thus, in this limit, according to eq.(\ref{40}),
\begin{equation}
\label{43}
\lim_{\beta\to 0} \; U_{\lambda}(z) \; \to \; \delta(z)
\end{equation}
so that eq.(\ref{38}) reduces to
\begin{equation}
\label{44}
\phi''(z) \; - \;  \phi(z) \; + \;  \phi^{2}(z) \; = \; 0 
\end{equation}
One can easily check (numerically) that this equation has an  instanton-like
solution with 
$\phi(0) \simeq 1.50 \; , \phi'(0) = 0$ and $\phi(z\to\infty) \to 0$
(see Fig.1)
\begin{figure}[h]
	\begin{center}
		\includegraphics[width=10.0cm]{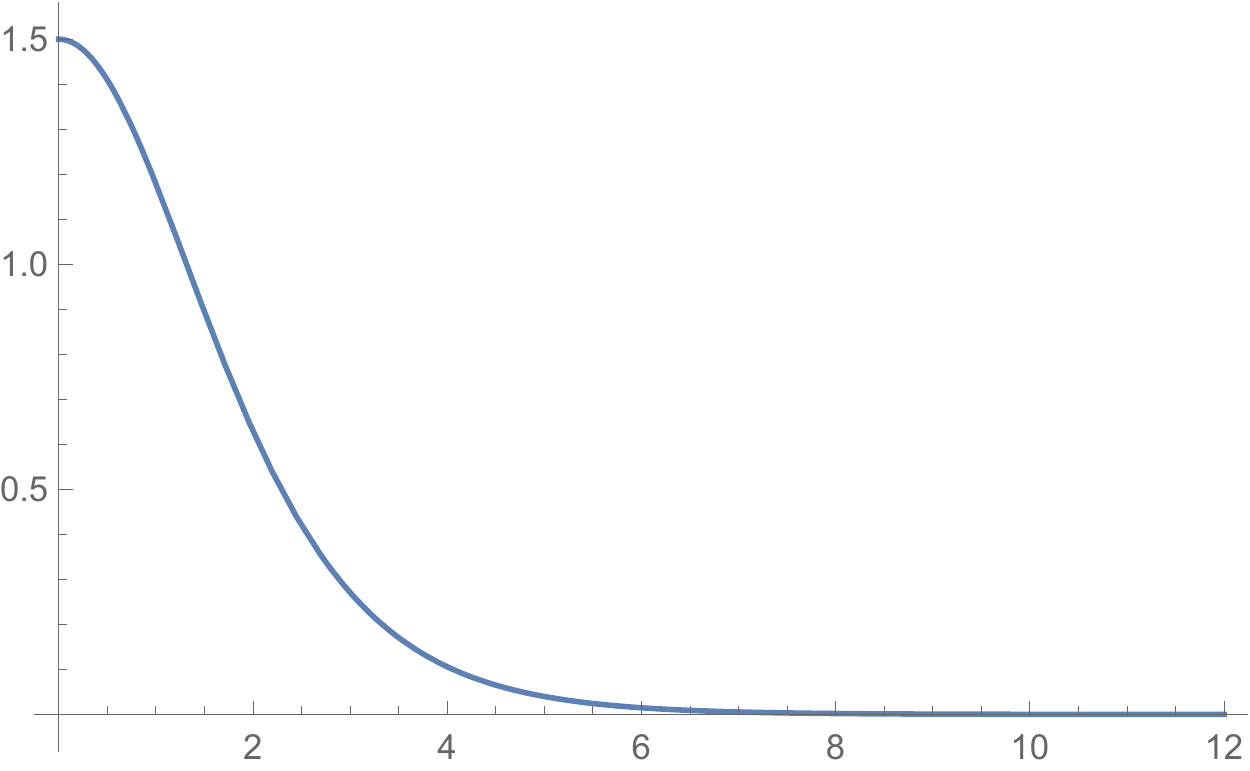}
		\caption[]{Instanton solution of eq.(\ref{44})}
	\end{center}
	\label{Figure1}
\end{figure}
Substituting this solution into eq.(\ref{42}) we find 
\begin{equation}
\label{45}
I \; = \; \int_{-\infty}^{+\infty} dz \, \phi(z) \; \simeq \; 6.00
\end{equation}
Thus, according to eqs.(\ref{27}), (\ref{28}), (\ref{41}) and (\ref{45}) we find
\begin{equation}
\label{46}
E_{N} \; \simeq \;  -\frac{1}{36} \beta^{5} \, u^{2} \, N^{3} \; \propto \; - N^{3} 
\end{equation}
This result, except for the numerical prefactor,  perfectly fits with 
the exact value of ground state energy $-\frac{1}{24} \beta^{5} \, u^{2} \, N^{3}$
of the one-dimensional $N$-particle boson system
(see e.g. \cite{Rev}) and correspondingly provide the well known value 
of the free energy scaling exponent $\theta = 1/3$.

\subsection{$(2+1)$ directed polymers}

The situation in the $(2+1)$ case is more complicated.
As there are no reasons to expect that
the ground state solution of the original Schr\"odinger equation (\ref{11})
is anisotropic, in what follows it will be assumed that the function
$\psi({\bf r})$ is radially symmetric:  
$\psi({\bf r}) = \psi(|{\bf r}|) \equiv \psi(r)$. 
In this case equations (\ref{32})-(\ref{33}) take the form
\begin{equation}
\label{47}
\psi''(r) \; + \; \frac{1}{r} \psi'(r) \; - \; \lambda \psi(r) \; + \; 
\kappa \, \psi(r) \,
\int d^{2} r' \, U_{0}(|{\bf r} - {\bf r}'|) \, \psi(r') \; = \; 0
\end{equation}
\begin{equation}
\label{48}
2\pi \int_{0}^{\infty} dr \, r \, \psi(r) \; = \; 1
\end{equation}
Redefining
\begin{equation}
\label{49}
\psi(r) \; = \; \frac{\lambda}{\kappa} \, \phi\bigl(\sqrt{\lambda} \, r)
\end{equation}
and denoting $\sqrt{\lambda} \, {\bf r} \; = \; {\bf z}$,
instead of eqs.(\ref{47})-(\ref{48}) we get
\begin{equation}
\label{50}
\phi''(z) \; + \; \frac{1}{z} \phi'(z) \; - \; \phi(z) \; + \; 
\phi(z) \,
\int d^{2} z' \, U_{\lambda}(|{\bf z} - {\bf z}'|) \, \phi(z') \; = \; 0
\end{equation}
\begin{equation}
\label{51}
2\pi \int_{0}^{+\infty} dz \, z \, \phi(z) \; = \; \kappa
\end{equation}
where
\begin{equation}
\label{52}
U_{\lambda}(|{\bf z}|) \; = \; \frac{1}{2\pi\lambda} \; 
\exp\biggl\{-\frac{1}{2\lambda} \, |{\bf z}|^{2} \biggr\}
\end{equation}
The main difference with the one-dimensional case is that now 
in the limit $\lambda \to 0$ the value of $\kappa$, eq.(\ref{51})
remains finite. Indeed, according to eq.(\ref{52}),
\begin{equation}
\label{53}
\lim_{\lambda\to 0} \, U_{\lambda}(|{\bf z}|) \; = \; \delta({\bf z})
\end{equation}
In this case eq.(\ref{50}) reduces to 
\begin{equation}
\label{54}
\phi''(z) \; + \; \frac{1}{z} \phi'(z) \; - \; \phi(z) \; + \; 
\phi^{2}(z) \; = \; 0
\end{equation}
\begin{figure}[h]
	\begin{center}
		\includegraphics[width=10.0cm]{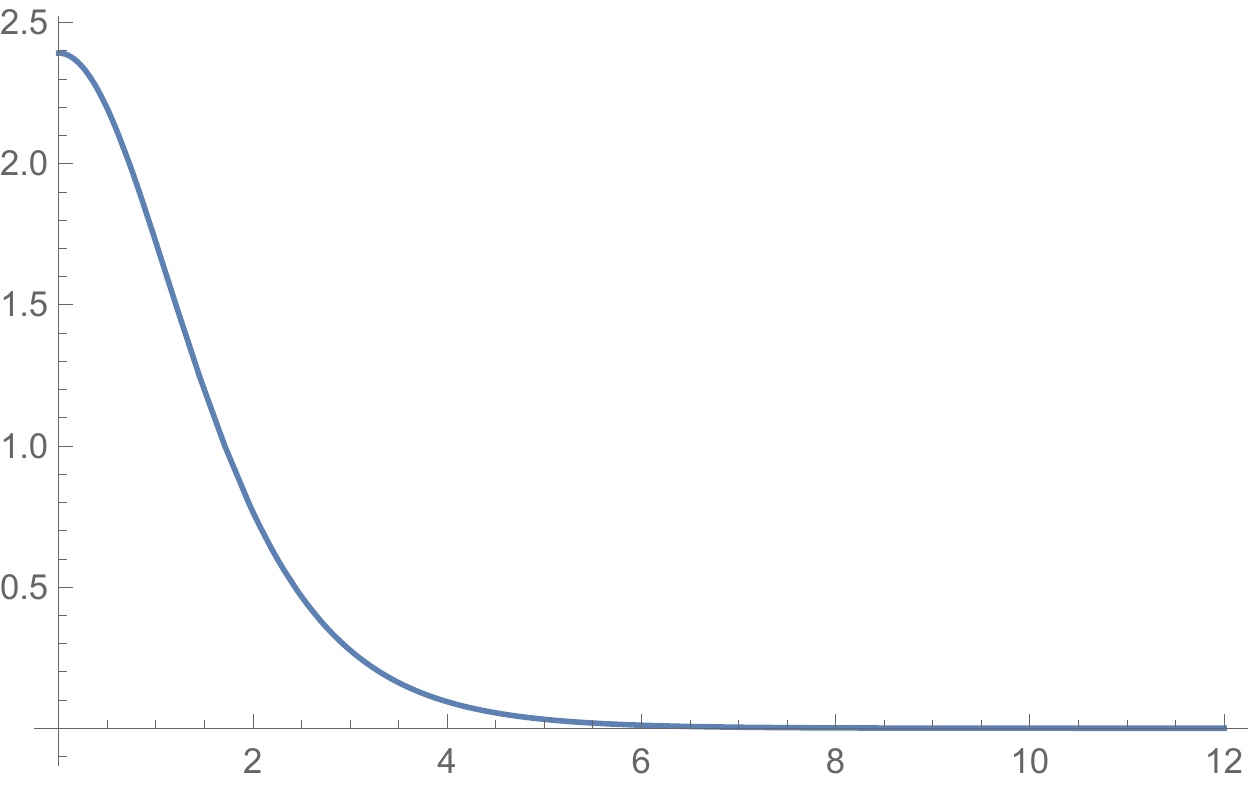}
		\caption[]{Instanton solution of eq.(\ref{54})}
	\end{center}
	\label{Figure2}
\end{figure}
This equation has an  instanton-like solution with 
$\phi(0) \simeq 2.39 \; , \phi'(0) = 0$ and $\phi(z\to\infty) \to 0$
(see Fig.2).
Substituting this solution into eq.(\ref{51}) we find that
at $\lambda=0$
\begin{equation}
\label{55}
\kappa(\lambda=0) \; \equiv \; \kappa_{0} \; \simeq \; 31.00
\end{equation}
At non-zero $\lambda \ll 1$, for $\kappa > \kappa_{0}$ 
numerical solution of eqs.(\ref{50})-(\ref{51}) 
demonstrate perfect linear dependence (see Fig.3)
\begin{equation}
\label{56}
\lambda(\kappa) \; = \; \gamma \, (\kappa - \kappa_{0})
\end{equation}
with
\begin{equation}
\label{57}
\gamma \; \simeq \; 0.050
\end{equation}
\begin{figure}[h]
	\begin{center}
		\includegraphics[width=10.0cm]{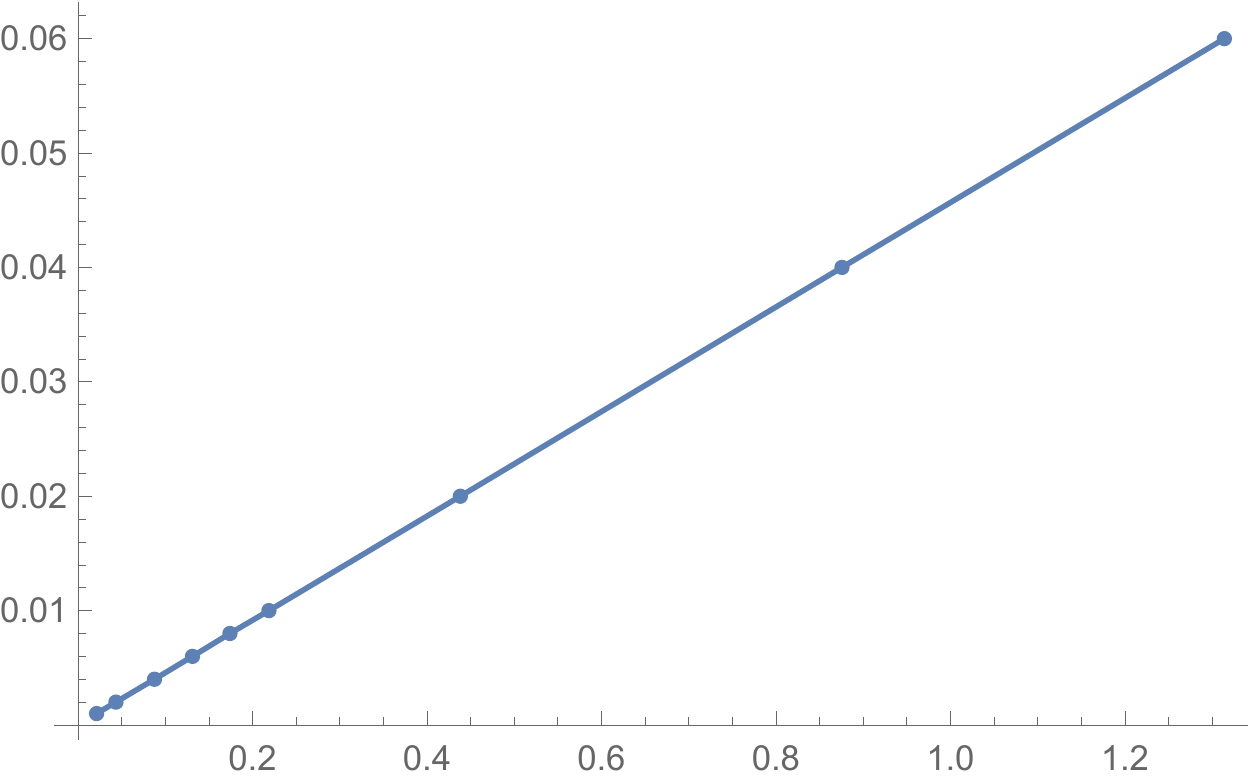}
		\caption[]{Dependence of $\lambda$ on $(\kappa-\kappa_{0})$, eq.(\ref{56})}
	\end{center}
	\label{Figure3}
\end{figure}
On the other hand, in the region $\kappa < \kappa_{0}$ 
by numerical methods we  find no non-negative solutions of equations
(\ref{50})-(\ref{51})
(such that $\psi({\bf r}\to\infty) \to 0$) for any value of $\lambda$.

%%%%%%%%%%%%%%%%%%%%%%%%%%%%%%%%%%%%%%%%%%%%%%%%%

\section{Free energy scaling }

Substituting eqs (\ref{22})  and (\ref{27})
into eq.(\ref{56}) for the mean-field ground state energy of 
considered two-dimensional $N$-particle boson system we find
\begin{equation}
\label{58}
E_{N} \; \simeq \; -\frac{\gamma}{4\beta \epsilon^{2}} \, N\, 
              \bigl(2\beta^{3} u N \; - \; \kappa_{0}\bigr)
\end{equation}
where $\epsilon$ is the correlation size of the random potential, eq.(\ref{3}),
and the numerical (approximate) values of the factors $\gamma$ and $\kappa_{0}$
are given in eqs.(\ref{55}) and (\ref{57}). Note that the above result is valid only
for 
\begin{equation}
\label{59}
N \; > \; N_{*} \, \equiv \, \frac{\kappa_{0}}{2\beta^{3} u}  \; \gg \; 1
\end{equation}
Thus, according to eqs.(\ref{58}), (\ref{8a}) and(\ref{10}), 
in the limit $t \to \infty$ for the replica partition function
we find the following estimate
\begin{equation}
\label{60}
Z(N,t) \; \sim \; \exp\biggl\{\frac{\gamma}{4\beta \epsilon^{2}} \, N\, 
\bigl(2\beta^{3} u N \; - \; \kappa_{0}\bigr) \, t\biggr\}
\end{equation}
Correspondingly, we see that in the high-temperature limit and at large $N$
 eq.(\ref{5}) reads
\begin{equation}
\label{61}
\int_{-\infty}^{+\infty} dF \, P(F) \, \exp\bigl\{-\beta N F\bigr\} \; \sim \; 
\exp\biggl\{\frac{\gamma u}{2 \epsilon^{2}} \,(\beta N)^{2} \, t
 \; - \; \frac{\gamma \kappa_{0}}{4\beta^{2} \epsilon^{2}} \, \beta N \, t\biggr\}
\end{equation}
As in the exponential in r.h.s of the above relation there is linear in 
$N$ term the total free energy $F$ can be redefined such that it splits 
into two {\it independent} parts: $F \, = \, \overline{F} \, + \, \tilde{F}$,
where $\overline{F}  =  \gamma \kappa_{0} t/(4\beta^{2} \epsilon^{2})$ 
is an extensive  non-random (selfaveraging) part while
$\tilde{F}$ is the fluctuating contribution described by a distribution function
$\tilde{P}\bigl(\tilde{F}\bigr)$ which according to eq.(\ref{61}) is defined
by the relation
\begin{equation}
\label{64}
\int_{-\infty}^{+\infty} d\tilde{F} \, \tilde{P}\bigl(\tilde{F}\bigr) 
\, \exp\bigl\{-\beta N \tilde{F}\bigr\} \; \sim \; 
\exp\biggl\{\frac{\gamma u}{2 \epsilon^{2}} \, t \, (\beta N)^{2} \biggr\}
\end{equation}
In the case the above relation would be valid for {\it any} $N$ we would
find that $\tilde{P}\bigl(\tilde{F}\bigr)$ is just simple Gaussian distribution
function. In fact, as eq.(\ref{64}) is valid only for $N \, > \, N_{*} \, \gg \, 1$,
eq.(\ref{59}), it gives us only the {\it left} tail of this distribution:
\begin{equation}
\label{65}
\tilde{P}\bigl(\tilde{F} \to -\infty\bigr) \; \sim \; 
\exp\biggl\{ -\frac{\epsilon^{2}}{2\gamma \, u \, t} \, \tilde{F}^{2}  \biggr\}
\end{equation}
Note that this asymptotics sets in at 
$|\tilde{F}| \gg (\beta N_{*}) \, \gamma u t/\epsilon^{2} \, \sim \, 
t/(\beta^{2}\epsilon^{2})$.

It would be natural to suppose that  
the entire (unknown) probability distribution $\tilde{P}(\tilde{F})$
is a {\it universal} function in a sense that it is defined by the only energy scale
so that it can be represented in the form 
$\tilde{P}(\tilde{F}) = G\bigl(\tilde{F}/F_{0}(t))$, where
$G(x)$ is some unknown universal function and $F_{0}(t)$ is the typical energy scale 
of the free energy fluctuations. 
It is supposed that this form holds
for {\it all} scales of $\tilde{F}$ including far tails where $|\tilde{F}| \gg F_{0}(t)$.
If the above hypothesis is correct then using the  explicit expression for far left tail 
of the distribution  function $\tilde{P}\bigl(\tilde{F}\bigr)$, eq.(\ref{65}),  
one immediately finds that in the considered 
high-temperature limit the typical value of the free energy fluctuations scale as
\begin{equation}
\label{66}
\tilde{F} \; \sim \; F_{0}(t) \; = \; \frac{\sqrt{\gamma \, u}}{\epsilon} \, t^{1/2}  
\end{equation}
where $\gamma \simeq 0.05$, eq.(\ref{57}), and 
$u$ and $\epsilon$ are the strength and the correlation length of the 
random potential, eqs.(\ref{2})-(\ref{3}).
The above eqs.(\ref{65})-(\ref{66}) constitute the main results of 
the present research.

\section{Conclusions}

Regardless of their somewhat "trivial" form, the results presented in 
this paper, eqs. (\ref{65}) and (\ref{66}), imply several
rather non-trivial conclusions.

First of all, it should be noted that the prefactor in the time scaling of the free energy fluctuations,  eq.(\ref{66}), is defined by the parameters
of the disorder potential. Thus 
the estimate, eq.(\ref{66}), is in a full agreement with the rigorous proof 
\cite{Lacoin,Berger-Lacoin,Comets} which states that even in the high temperature
limit the statistical properties of (2+1) directed polymers are defined by
the random potential. 

Second,  
the fact that in the high temperature limit the free energy time 
scaling exponent $\theta = 1/2$ is different from the one at the zero
temperature (which is close to $0.241$) means that in two dimensions this
scaling exponent is temperature dependent. 
Maybe this takes place because
the dimension $d=2$ is critical for ($d$ + 1) directed polymers.

Third, it is also interesting to note that in the high-temperature limit 
the prefactor in time scaling of the free energy fluctuations,
eq.(\ref{66}), turns out to be temperature independent 
(unlike the one-dimensional case, where it is proportional to $\beta^{2/3}$).

	It should be stressed however, that all the results presented in this paper 
	are based on two crucial assumptions.  
	The first one is pure heuristic mean-field ansatz, eq.(\ref{25}). 
	It looks quite reasonable in the limit of large number of replicas (which according to eq.(\ref{59}) corresponds to the high-temperature limit). Moreover,
	it works very well in the one-dimensional case (see Section III.A). 
	The second one is the hypothesis that in the considered system 
	the entire  probability distribution function of the free energy fluctuations
	$\tilde{P}(\tilde{F})$ reduces to  a {\it universal} function: 
	$\tilde{P}(\tilde{F}) = G\bigl(\tilde{F}/F_{0}(t))$ and this is valid 
	for {\it all} energy scales including far tails where $|\tilde{F}| \gg F_{0}(t)$. 
	For the moment the only support for this hypothesis is that {\it usually}
	in physical systems exhibiting scaling phenomena  it is correct. In particular, it is 
	certainly correct for (1+1) polymers where the exact solution provide us with
	the universal function $G(x)$ which is the Tracy-Widom distribution function.
	On the other hand, although both assumptions looks quite reasonable this 
	does not  guarantee that they are correct...
	In view of that, further analytic studies as well as numerical simulations of the
	considered system at {\it finite} (or high) temperatures would be extremely helpful.

\acknowledgments

VD is thankful to Quentin Berger for fruitful discussion of mathematical and 
probabilistic aspects of (d+1) directed polymers problem.
VD is also grateful to Pierre Le Doussal, Jeremy Quastel, Lev Ioffe, Sergei Nechaev and Maxim Dolgushev for numerous illuminating discussions.

%%%%%%%%%%%%%%%%%%%%%%%%%%%%%%%%%%%%%%%%%


\begin{thebibliography}{99}
	
%%%%%%%%%%%%% Introduction

\bibitem{KPZ} 
Kardar M, Parisi G, Zhang Y-C,
Phys.\ Rev.\ Lett.\ {\bf 56}, 889 (1986)

\bibitem{Corwin} 
Corwin I,
Random Matrices: Theory Appl. {\bf 1}, 1130001 (2012)

\bibitem{Borodin} 
Borodin A,  Corwin I and Ferrari P,
Comm. Pure Appl. Math. {\bf 67}, 1129–1214 (2014)

\bibitem{Rev} 
V.Dotsenko 
{\it Statistical properties of one-dimensional directed polymers in a random
	potential}, arXiv:1703.04305 (2017)

%%%%%%%%%%%%%%

\bibitem{numerics1}  
Halpin-Healy  T, 
Phys.\ Rev.\ Lett.\ {\bf 109}, 170602 (2012)

\bibitem{numerics2}  
Halpin-Healy  T, 
Phys.\ Rev. \ E {\bf 109}, 042118 (2013)

\bibitem{numerics3}  
Halpin-Healy  T and  Palasantzas G, 
EPL, {\bf 105}, 50001 (2013)

%%%%%%%%%%%%%%%%%%%%%%%%%%%


\bibitem{Lacoin}
Lacoin H, 
Commun. Math. Phys. {\bf 294} 471 (2020)

\bibitem{Berger-Lacoin}
Berger Q and Lacoin H,
Ann. Inst. Henri Poincaré: Probab. Stat. {\bf 53} 430 (2017)

\bibitem{Comets}
Comets F,
{\it Lecture notes from the 46th Probability Summer School, Saint-Flour, 2016}, 
Springer International Publishing (2017)

\bibitem{Bolthausen}
Bolthausen E,
Commun. Math. Phys. {\bf 123} 529 (1989)


\bibitem{Albert-Khanin-Quastel-1}
Alberts T, Khanin K and Quastel J,
Phys. Rev. Lett. {\bf 105} 090603 (2010)
	
\bibitem{Albert-Khanin-Quastel-2}
Alberts T, Khanin K and Quastel J,
Ann. Probab. {\bf 43} 1212 (2014)

%%%%%%%%%%%%%%%%%%%%%%%%%%%

\bibitem{hhf_85} Huse D A , Henley C L  and Fisher D S, 
Phys.\ Rev.\ Lett.\ {\bf 55}, 2924 (1985)

\bibitem{numer1} Huse D A and Henley C L, 
Phys.\ Rev.\ Lett. {\bf 54}, 2708 (1985)

\bibitem{numer2} Kardar M and Zhang Y-C, 
Phys.\ Rev.\ Lett.\  {\bf 58}, 2087 (1987)


\bibitem{kardar_87} Kardar M, 
Nucl.\ Phys.\ {\bf B 290}, 582  (1987)


\bibitem{bouchaud-orland} Bouchaud J P and Orland H, 
J.\ Stat.\ Phys.\ {\bf 61}, 877 (1990)

\bibitem{hh_zhang_95} Halpin-Healy  T and Zhang Y-C,
Phys.\ Rep.\ {\bf 254}, 215 (1995)




\bibitem{Johansson}  Johansson K, 
Comm. \ Math. \ Phys. \ {\bf 209}, 437 (2000)

\bibitem{Prahofer-Spohn}  Prahofer M and Spohn H, 
J.\ Stat.\ Phys.\ {\bf 108}, 1071 (2002)

\bibitem{Ferrari-Spohn1} Ferrari P L and Spohn H, 
Comm. \ Math. \ Phys. \ {\bf 265}, 1 (2006)



\bibitem{KPZ-TW1a} Sasamoto T and Spohn H, 
Phys.\ Rev.\ Lett.\ {\bf 104}, 230602 (2010)

\bibitem{KPZ-TW1b} Sasamoto T and Spohn H, 
Nucl.\ Phys.\ {\bf B834}, 523 (2010)

\bibitem{KPZ-TW1c}Sasamoto T and Spohn H,
J.\ Stat.\ Phys. {\bf 140}, 209 (2010)


\bibitem{KPZ-TW2}  Amir G , Corwin I and Quastel J, 
Comm.\ Pure Appl.\ Math.\ {\bf 64}, 466 (2011)


\bibitem{LeDoussal1} Calabrese P, Le Doussal P and Rosso A,
EPL, {\bf 90}, 20002  (2010)

\bibitem{BA-TW2} Dotsenko V, 
EPL, {\bf 90}, 20003 (2010)

\bibitem{BA-TW3}  Dotsenko V, 
J.Stat.Mech. P07010 (2010)


\bibitem{LeDoussal2} Calabrese P and  Le Doussal P, 
Phys.\ Rev.\ Lett.\ {\bf 106}, 250603  (2011)


\bibitem{goe}  Dotsenko V,
J. Stat. Mech. P11014  (2012)

\bibitem{end-point} 
Dotsenko V, 
J. Stat. Mech.,  P02012  (2013)


\bibitem{LeDoussal3} 
Gueudr\'e T and  Le Doussal P, 
EPL, {\bf 100}, 26006  (2012)

%%%%%%%%%%%%%%%%

\bibitem{TW-GUE} 
Tracy C A  and Widom H, 
Commun.Math Phys., {\bf 159}, 151 (1994)


%%%%%%%%%%%%%%%%%%%%%%%%%%%%%%%%%%%%%%%%%%%%%%




\end{thebibliography}
\end{document}